\documentclass[cmpj,superscriptaddress,showkeys,twocolumn]{revtex4-2}
\usepackage{graphicx}
\usepackage{dcolumn}
\usepackage{bm}
\usepackage[utf8]{inputenc}
\usepackage{float}
\usepackage{color}
\definecolor{rojo}{rgb}{1,0,0}
\definecolor{verde}{rgb}{0,0.8,0.5}
\definecolor{azul}{rgb}{0,0,1}
\definecolor{rosa}{cmyk}{0,1,0,0}
\usepackage{bm}
\usepackage{color}

\begin{document}

\title{Wavelength-Selective control of Atomic Scale Au Contacts}

\author{Werner Br\"amer-Escamilla}
\email{wbramer@yachaytech.edu.ec}
\affiliation{School of Physical Sciences and Nanotechnology, Yachay Tech University, Urcuquí 100119, Ecuador}

\author{Floralba Lopez}
\affiliation{School of Chemical Sciences and Engineering, Yachay Tech University, Urcuquí 100119, Ecuador}

\author{Laila Procel}
\affiliation{School of Physical Sciences and Nanotechnology, Yachay Tech University, Urcuquí 100119, Ecuador}

\author{David Llerena}
\affiliation{Departamento de F\'isica, Colegio de Ciencias e Ingenier\'ia, Universidad San Francisco de Quito, Diego de Robles y V\'ia Interoce\'anica, Quito, 170901, Ecuador}

\author{Carlos Sabater}
\affiliation{Departamento de Física and Instituto Universitario de Materiales de Alicante (IUMA), Universidad de Alicante, E-03690 Alicante, Spain.}

\author{Ernesto Medina}
\email{emedina@usfq.edu.ec}
\affiliation{Departamento de F\'isica, Colegio de Ciencias e Ingenier\'ia, Universidad San Francisco de Quito, Diego de Robles y V\'ia Interoce\'anica, Quito, 170901, Ecuador}
\affiliation{School of Molecular Sciences, Arizona State University, 551E University Dr, Tempe, AZ 85281, USA}
\date{\today}

\begin{abstract}
We demonstrate wavelength-selective control of atomic motion in a mechanically controllable Au break junction. 
Excitation at $\lambda_{\mathrm{form}}\simeq 530~{\rm nm}$ drives gap closure and metallic bridge formation, whereas excitation at $\lambda_{\mathrm{rup}}\simeq 407~{\rm nm}$ drives neck thinning, bridge rupture, and subsequent gap opening. Unlike conventional optical switching in metallic contacts, where illumination primarily acts via thermal expansion, the present experiment reveals oppositely directed atomic drift at different wavelengths. Time-resolved conductance traces allow us to distinguish two dynamical regimes. In the tunneling regime, exponential conductance transients measure the drift velocity of the gap coordinate for both gap closure and gap opening. In the metallic regime, the Sharvin relation converts linear $\sqrt{G/G_0}$ transients into radial neck-growth and neck-thinning velocities of comparable magnitude. These results establish optically selected atomic drift as a mechanism for reversible control of metallic nanocontacts and provide a quantitative route to follow plasmon-assisted atomic rearrangements in real time.

\end{abstract}

\maketitle

\section{Introduction}

Light--matter interaction in metallic nanostructures can concentrate
optical energy into regions far smaller than the incident wavelength. In particular, localized plasmonic excitations produce intense near fields in metallic nanogaps and sharp constrictions
\cite{ward2010l,herzog2013dark}. The resulting optical response is highly
sensitive to the local atomic geometry: small variations in the gap
width or in the arrangement of atoms at a metallic contact can
substantially modify both the plasmonic modes and the electronic
conductance. This strong coupling between optical response, electronic
transport, and atomic structure creates the possibility of using light
not only to probe metallic nanojunctions, but also to control their
formation and rupture \cite{benz2016}.

Atomic-scale metallic junctions provide a particularly direct platform for studying this coupling. Such junctions are commonly produced using scanning-tunneling-microscope break junctions or mechanically controllable break junctions
\cite{Pascual1993,Krans93,Krans96}. Their conductance is extremely sensitive to the atomic configuration of the contact. When the electrodes are joined by an atomic-scale metallic bridge,
electronic transport is described by the Landauer expression
$G=G_0\sum_i T_i$
where $G_0=2e^2/h$ is the conductance quantum and $T_i$ is the
transmission probability of the $i$th electronic channel
\cite{Landauer57,Cuevasbook}. For a wider ballistic contact
supporting many transverse modes, this description approaches the
semiclassical Sharvin relation,
$G\simeq G_0\frac{k_F^2A}{4\pi}$
where $k_F$ is the Fermi wavevector and $A$ is the minimum
cross-sectional area of the metallic neck \cite{sharvin1965}. By
contrast, when the electrodes are separated by a nanometric gap, the
conductance is governed by tunneling and depends exponentially on the electrode separation $d$,
$G(d)=G_{\rm pref}\exp(-\beta d)$,
where $\beta$ is an effective tunneling decay constant. These distinct
conductance laws make time-resolved transport measurements a sensitive
probe of both the opening and closing of a nanogap and the growth or
thinning of a metallic neck.

Previous work has shown that illumination can reversibly switch bare
metallic atomic contacts without requiring molecular photochromic
units. In particular, Zhang \textit{et al.} demonstrated optically
controlled switching between tunneling and ballistic-contact regimes
and attributed the effect primarily to plasmonic heating and the
resulting thermal expansion of the electrodes \cite{Zhang2019}.
Plasmonic excitation may, however, influence atomic motion through
additional mechanisms. Depending on the optical frequency and junction
geometry, illumination can generate hot carriers, local temperature
gradients, enhanced surface mobility, and optical near-field forces
\cite{Marchesin2015,Khurgin2024,Zhou2024}. It therefore remains an open
question whether optical excitation can do more than increase the
temperature or expand a metallic junction. In particular, it is not
known whether different excitation wavelengths can bias the atomic
dynamics in opposite directions, driving the same type of
nanocontact either toward bridge formation or toward rupture.

Here we demonstrate wavelength-selective atomic motion in a
mechanically controllable Au break junction under ambient conditions.
At mechanically fixed operating points, illumination at the formation
wavelength, $\lambda_{\rm form}\simeq 530\,{\rm nm}$,
drives the closure of an initially open nanogap and promotes the growth
of a metallic bridge. Illumination at the rupture wavelength,
$\lambda_{\rm rup}\simeq 407\,{\rm nm}$,
instead drives the thinning of an existing metallic neck, its rupture, and the subsequent opening of the gap. Thus, the two wavelengths select opposite directions of atomic rearrangement.

Time-resolved conductance measurements allow us to quantify the
dynamics on both sides of the contact transition. In the tunneling
regime, exponential conductance transients provide the velocity of the
gap coordinate during both gap closure and gap opening. In the
many-channel metallic regime, the Sharvin relation converts the time
dependence of $\sqrt{G/G_0}$ into the radial growth or thinning velocity
of the metallic neck. The velocities obtained from these two
independent descriptions are comparable, supporting a continuous
drift-like evolution across the transition between a metallic bridge
and an open gap. The dark relaxation of the junction after the
illumination is removed further indicates that light supplies a
wavelength-dependent driving term that competes with the mechanical
bias of the junction.

To interpret the reversal of the atomic motion, we introduce a
phenomenological wavelength-dependent effective-energy landscape in
which illumination modifies the relative stability and activation
barriers of bridged and ruptured configurations. This description is
not intended as a microscopic theory of the optical forces, but as a
framework for distinguishing wavelength-selected atomic drift from a
purely scalar heating mechanism. The results establish a transport-based method for following plasmon-assisted atomic rearrangements in real time and demonstrate reversible optical control of metallic
nanocontacts by selecting the excitation wavelength.

The remainder of the paper is organized as follows. Section~II
describes the experimental setup and switching protocols. Section~III analyzes the light-driven dynamics in the tunneling and metallic regimes, as well as the dark relaxation of the junction.
Section~IV introduces a phenomenological wavelength-dependent
effective-energy landscape, and Sec.~V summarizes the main results
and their implications.

\begin{figure}
		\centering
		\includegraphics[scale=0.4]{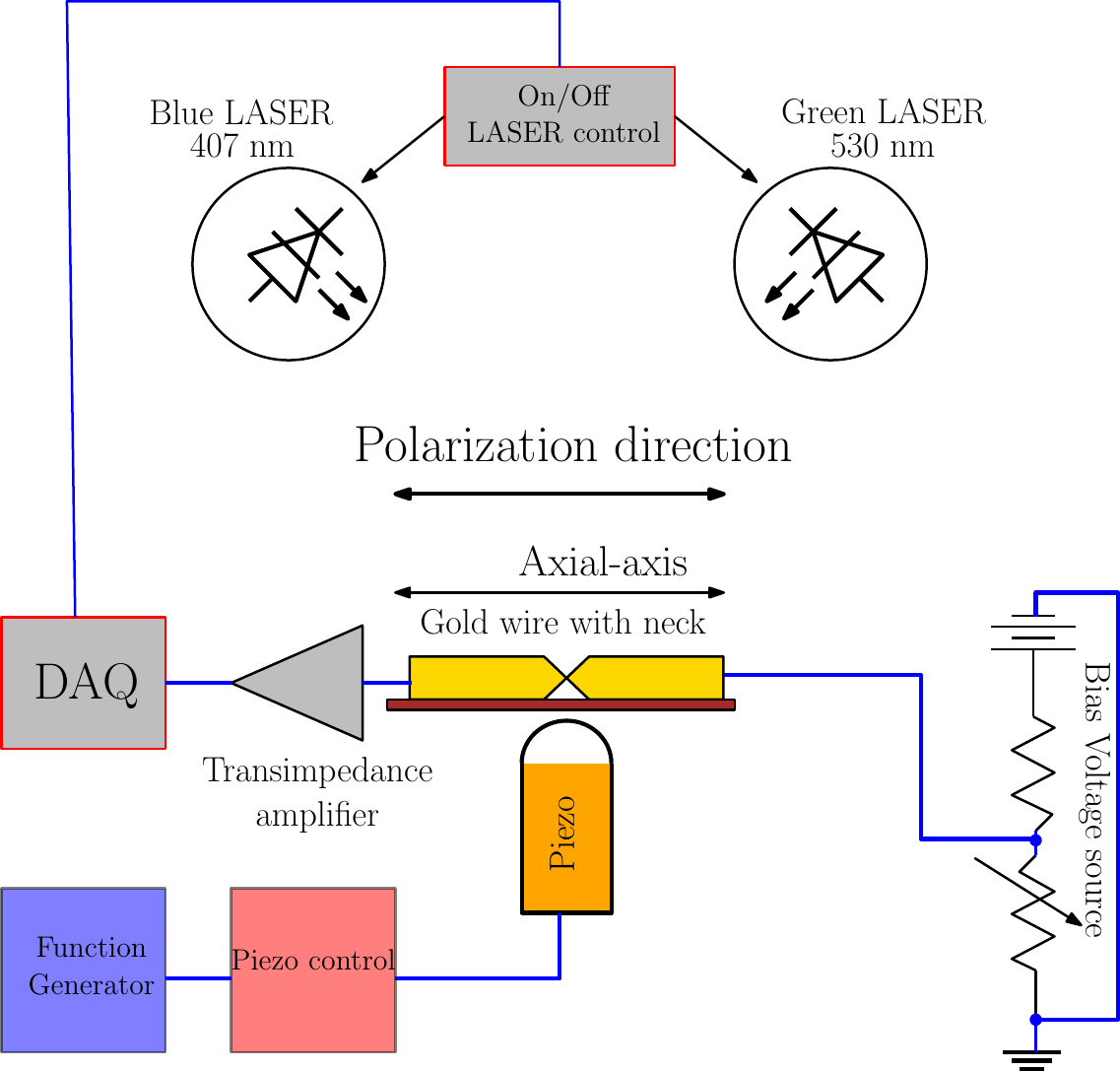}
		\caption{Diagrammatic representation of the experimental setup. The distance between the contacts on the neck of the gold wire is controlled by a piezoelectric element, which is connected to a function generator via a piezoelectric controller. Both the conductance measurements and the switching of the lasers on and off are controlled by a computer via a DAQ (Data Acquisition Unit).}
		\label{fig:esquemadelequipo2}
	\end{figure}
    
\begin{figure*}[t]
    \centering
    \includegraphics[width=\textwidth]{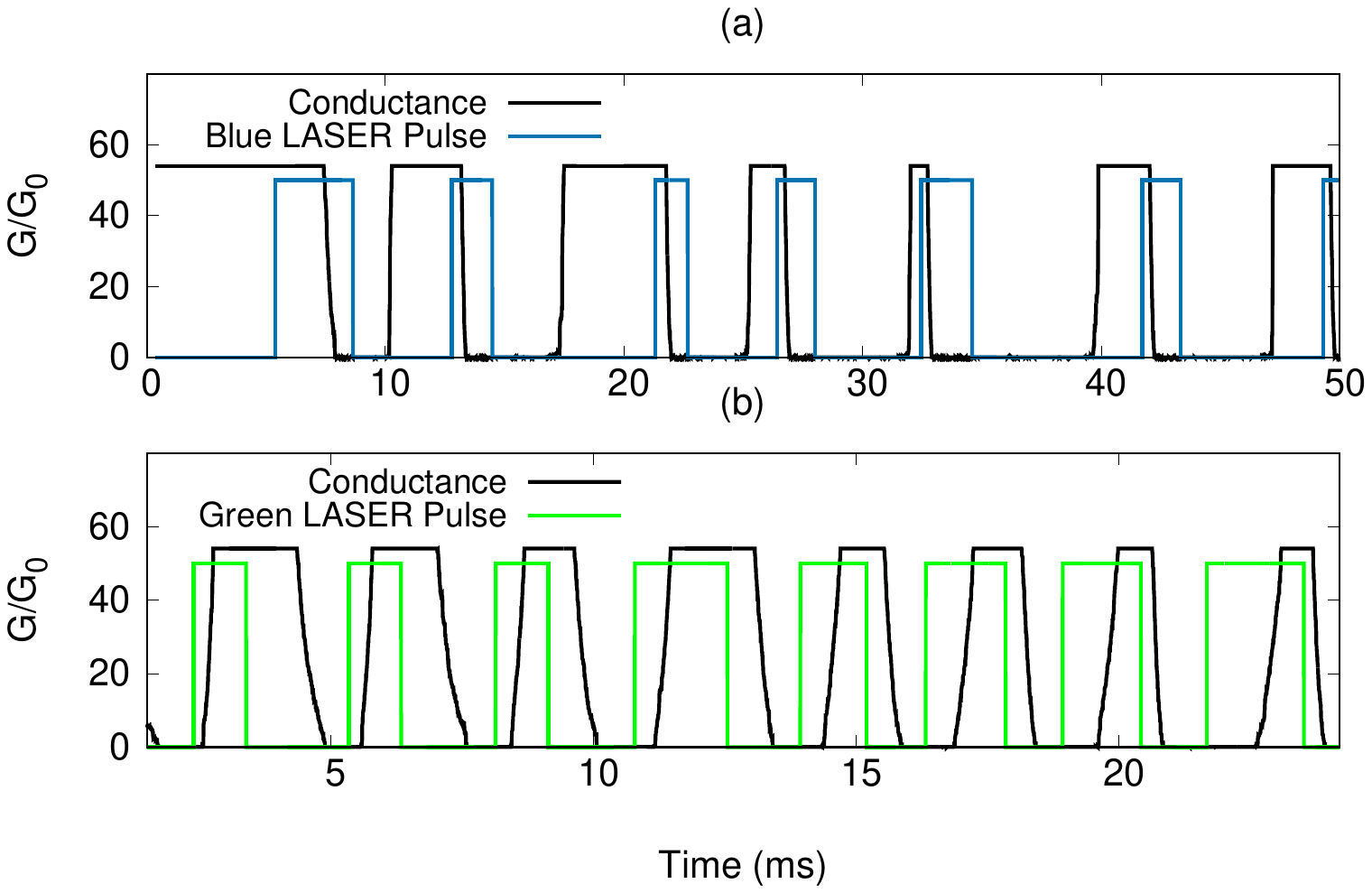}
    \caption{Experimental results for the conductance of a) an initially formed metallic bridge that is ruptured by the {\it blue} wavelength laser and b) an initially open gap, in which a metallic bridge is formed by the {\it green} wavelength laser. The black sequence of square pulses shows the junction conductance, while the blue (online) and yellow (online) represent laser pulses on the junction.}
    \label{fig:Multiple-pulse}
\end{figure*}

\section{Materials and methods}
	
	\subsection{Experimental setup}
		The mechanically controllable break junction (MCBJ) technique is used to perform the experiment under ambient conditions. The diagram of the setup is shown in Figure \ref{fig:esquemadelequipo2}. A high-purity gold wire (4N MaTeck), with a diameter of 300 $\mu$m and a neck at its center, is secured to a plastic substrate with epoxy adhesive. The MCBJ has two component $Z$ motions: the coarse $Z$-axis movement is controlled by a micrometer screw, and the fine movement is controlled by a piezoelectric actuator. The piezoelectric actuator is connected to a piezo controller (Kinesis from THORLABS), and the signal is produced by a function generator (BK PRECISION 4065). Measurements were taken with eight different bias voltages (60 mV, 105 mV, 151 mV, 202 mV, 252 mV, 314 mV, 359 mV, 410 mV). The switching phenomenon persisted for all bias voltages; the data shown are with a 202 mV bias. The gold wire is connected to a voltage source with a bias voltage fixed at 202 mV, and the output signal is fed to a transimpedance amplifier. Finally, the amplified signal goes to a Data Acquisition (DAQ) device (Measurement Computing USB-1208H-2AO). The DAQ also independently turns two lasers on and off using a control with an optocoupler isolator to avoid electrical noise. A computer takes measurements and controls the lasers' on/off state. The piezoelectric actuator's movement is manually controlled via the function generator's panel controls.  
	
	The light sources were two lasers, each with less than 300 mW: one green (530 nm) and one blue (407 nm). The elliptical polarization was converted to linear using a $\lambda/4$ plate and a linear polarizer with an extinction ratio of 9000:1. The electric field was oriented along the gold axial wire (along the wire's length).
	
	The first procedure was to use a micrometer screw to break the neck. Then the micrometer screw is pulled back to re-establish electrical contact. 
	
	\subsubsection{Bridge Rupture}
	
	 Using the function generator, the piezoelectric actuator is extended until the conduction measurement drops below the I-V saturation limit. This gold wire arrangement must remain stable over time; that is, it should not change within 24 hours. At this moment, the laser light source is turned on, which destroys the contact. Then it is turned off, the contact is restored on its own, and the cycle repeats about 20 times. 
	\subsubsection{Bridge formation}
	
	In this case, the piezoelectric actuator is extended until no conduction is detected, and the gold wire arrangement must remain stable over time as before. At this moment, the laser light source is turned on, forming the bridge; then it is turned off, and the bridge ruptures on its own, and the cycle is repeated about 20 times. 

\section{Results}


Figure \ref{fig:Multiple-pulse} shows the conductance dynamics for light-induced bridge rupture and bridge formation. In both protocols, up to 20 laser pulses are applied sequentially. Panel~(a) shows that blue illumination ruptures a pre-existing metallic bridge: the conductance drops when the blue laser is turned on, and the bridge reforms after the laser is turned off. Panel~(b) shows the complementary process: starting from an open gap, green illumination forms a metallic bridge, which ruptures again after the laser is turned off. Both behaviors are reproducible over many switching cycles.


We expect two regimes of conduction for either the rupture or the formation of the contact: a tunneling regime before bridge formation or right after bridge rupture, and a Sharvin regime when the contact evolves, either after it is already bridged after formation or during bridge rupture before the gap opens. 

\begin{figure}
    \centering
    \includegraphics[width=1.1\linewidth]{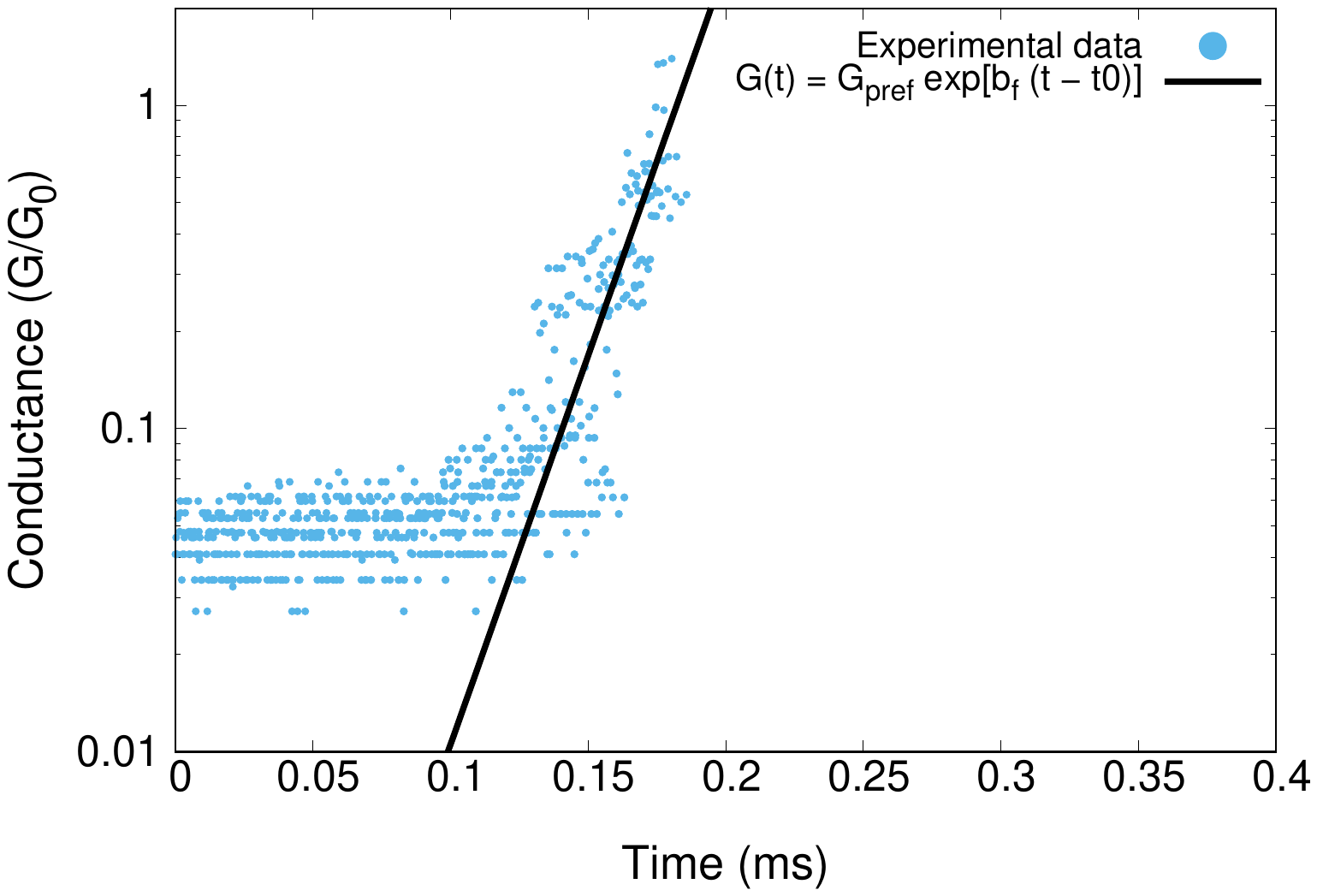}
    \caption{
Conductance below $G_0$ during gap closure prior to bridge formation under the formation wavelength.
The semilogarithmic behavior is well described by an exponential transient, $G(t)=G_{\rm pref}\exp[b_f(t-t_0)]$.
The physically relevant fit parameter is the logarithmic slope $b_f=55.46~{\rm ms}^{-1}$.
.}
    \label{fig:Tunnelclosing}
\end{figure}

Figure \ref{fig:Tunnelclosing} depicts the tunneling range for bridge formation. 
We adopt the following scenario for the time dependence of gap closure before bridge formation: In the tunneling regime, the two Au electrodes are separated by a nanometric gap, so the optical response can be qualitatively described in terms of coupled plasmonic modes of the two facing metallic asperities. When the incident field is polarized along the junction axis, charge oscillations on the two sides of the gap can hybridize into modes with distinct charge configurations \cite{Nordlander2004}. The mode relevant for contact formation is the one in which opposite induced charges face each other across the gap. This produces a large electric field in the gap and an attractive near-field optical force between the electrodes. As shown in Fig.\ref{fig:Multiple-pulse}, the 202 mV bias does not produce a sufficient field to move opposite sides of the contacts together.

\begin{figure}
    \centering
    \includegraphics[width=0.9\linewidth]{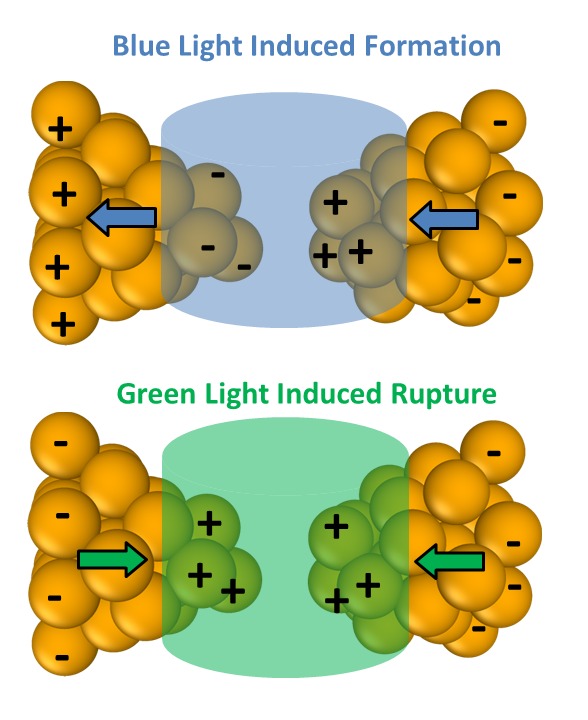}
    \caption{Schematic representation of wavelength-dependent gap modes once a nanogap is present. The rupture wavelength is associated with a destabilizing optical force that favors gap opening, whereas the formation wavelength is associated with an attractive optical force that favors gap closure. The schematic is not intended to describe the metallic-bridge stage, where the ordinary capacitive gap mode is quenched by charge flow through the bridge.}
    \label{fig:BondingAntibonding}
\end{figure}

The experimentally observed formation wavelength is interpreted as the wavelength that couples most efficiently to the attractive gap mode of the actual junction, rather than as a universal ``blue'' bonding plasmon. An illustration of this scenario is depicted on the right panel of Fig.\ref{fig:BondingAntibonding}.
Once a metallic bridge is formed, this gap-plasmon description no longer applies, since charge can flow through the bridge and the gap mode is quenched. The subsequent growth of the metallic contact is then better described by the Sharvin conductance and post-contact electromigration-assisted atomic drift, which we will address later.

\subsection{Bridge formation under the formation wavelength: drift-like gap closure}

We first consider the tunneling regime preceding bridge formation. In this regime, the two Au electrodes are separated by a nanometric gap, and the conductance is exponentially sensitive to the electrode separation $d$,
\begin{equation}
    G(d)=G_{\rm pref}e^{-\beta d},
\end{equation}
where $\beta=\frac{2\sqrt{2m\phi}}{\hbar}$
is the full tunneling decay constant, $m$ is the electron mass, and $\phi$ is an effective barrier height. For effective barriers in the range $\phi\simeq 3\text{--}5~{\rm eV}$, one obtains
$\beta\simeq1.77\text{--}2.29~{\rm \AA}^{-1}$. 

Taking the natural logarithm gives
\begin{equation}
    \ln G(t)=\ln G_{\rm pref}-\beta d(t).
\end{equation}
Thus, in the tunneling regime, the time dependence of $\ln G$ directly measures the time dependence of the gap coordinate.

The conductance shown in Fig.~3 is well described by an exponential growth,
\begin{equation}
G(t)=G_{\rm pref} \exp\left[b_f(t-t_0)\right],
\end{equation}
so that
\begin{equation}
    \frac{d\ln G}{dt}=b_f.
\end{equation}
and $t_0$ is the reference time where the fitted behavior begins. The fitted value is
$b_f=55.46~{\rm ms}^{-1}$.

This exponential increase indicates drift-like gap closure. Writing
\begin{equation}
    d_f(t)=d_{f,0}-v_f(t-t_0),
\end{equation}
one obtains
\begin{equation}
    \ln G(t)=C_f+\beta v_f(t-t_0),
\end{equation}
where $C_f=\ln G_{\rm pref}-\beta d_{f,0}$. Therefore,
$b_f=\beta v_f$, and the gap-closing velocity is
$v_f={b_f}/{\beta}$.
Using $b_f=55.46~{\rm ms}^{-1}$ and
$\beta\simeq1.77\text{--}2.29~{\rm \AA}^{-1}$, we obtain
$v_f\simeq24.2\text{--}31.3~{\rm \AA/ms}
\simeq2.4\text{--}3.1~\mu{\rm m/s}$.
For an intermediate barrier of $4~{\rm eV}$, corresponding to
$\beta\simeq2.05~{\rm \AA}^{-1}$, this gives
$v_f\simeq2.7~\mu{\rm m/s}$.

The exponential conductance transient, therefore, corresponds to a nearly constant closing velocity of the gap coordinate. We interpret this as drift-like gap closure driven by the formation wavelength, through a combination of optical forces, local heating, enhanced surface mobility, and electron-wind forces\cite{HoffmannVogel2017}. The extracted velocity is larger than drift velocities reported for extended Au nanobars, but the comparison is not direct: the present system involves an atomic-scale, current-crowded, optically assisted break junction. The value is consistent with a surface-dominated atomic rearrangement process localized at the nanoconstriction.

\subsection{Bridge rupture under the blue wavelength: drift-like gap opening}

We now consider the tunneling regime reached immediately after the blue wavelength ruptures the metallic bridge. In this regime, the two Au electrodes are separated by a nanometric gap, and the conductance depends exponentially on the electrode separation, $G(d)=G_{\rm pref}e^{-\beta d}$, as in the previous case.
The experimental trace in Fig.~5 is well described by an exponential decay,
\begin{equation}
G(t)=G_{\rm pref} \exp[-b_b(t-t_0)],    
\end{equation}
with $b_b=119.5~{\rm ms^{-1}}$.
Equivalently,
\begin{equation}
    \frac{d\ln G}{dt}=-b_b.
\end{equation}
This behavior indicates a nearly constant gap-opening velocity rather than a diffusion-limited square-root law. Writing $d_b(t)=d_{b,0}+v_b(t-t_0)$,
one obtains
\begin{equation}
    \ln G(t)=C_b-\beta v_b(t-t_0),
\end{equation}
and therefore $v_b={b_b}/{\beta}$. Using $\beta\simeq1.77\text{--}2.29~{\rm \AA^{-1}}$, the fitted value $b_b=119.5~{\rm ms^{-1}}$ gives
$v_b\simeq 52\text{--}68~{\rm \AA/ms}
\simeq 5.2\text{--}6.8~\mu{\rm m/s}$.
For an intermediate barrier of (4~{\rm eV}), this corresponds to $v_b\simeq 5.8~\mu{\rm m/s}$.

This value is close to the radial thinning velocity extracted independently from the Sharvin regime under blue illumination, $|v_r|\simeq 4.6\text{--}5.6~\mu{\rm m/s}$. The agreement suggests that the rupture dynamics are governed by a directed drift process in both regimes: first by thinning of the metallic neck while $G>G_0$, and then by continued separation of the electrodes after the conductance enters the tunneling regime. The latter stage can be interpreted as optically assisted gap opening, driven by a combination of local heating, enhanced surface mobility, and a gap-mode optical force that favors separation once the metallic bridge has ruptured.

This interpretation is consistent with previous work on Au nanojunctions, in which voltage-induced heating and electromigration can thin gold contacts to the few-channel regime.\cite{Trouwborst2008} The velocity extracted here is much larger than drift velocities reported in extended Au nanobars,\cite{Singh}, but the comparison is not direct: the present system involves an atomic-scale break junction with strong current crowding, surface-dominated atomic motion, and optical excitation localized at the constriction. We therefore interpret $v_b$ as an effective drift velocity for an optically assisted atomic-scale rupture process, not as a bulk electromigration velocity.

\begin{figure}
    \centering
    \includegraphics[width=1.1\linewidth]{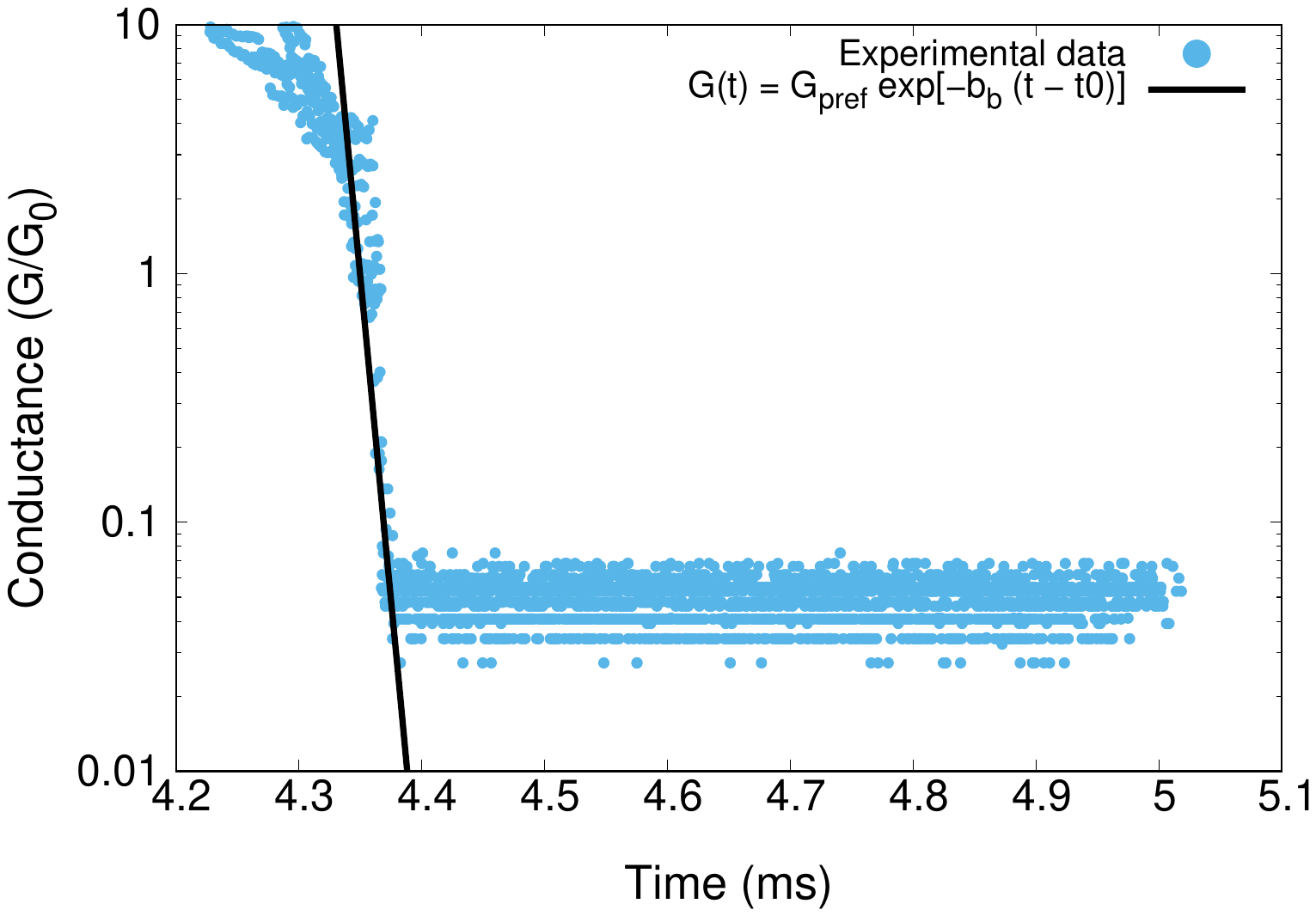}
    \caption{Conductance below $G_0$ after blue-light-induced bridge rupture. The exponential decay corresponds to drift-like gap opening in the tunneling regime. The physically relevant fit parameter is the logarithmic slope $b_b =119.5~{\rm ms}^{-1}$.}
    \label{fig:Tunnelopening}
\end{figure}

\subsection{Metallic nanocontact regime: Sharvin conductance and electromigration-assisted growth}

For a substantial fraction of the closing process, the conductance evolves through values ranging from a few conductance quanta to several tens of \(G_0\). In this regime, the two electrodes are connected through a contiguous metallic bridge, and transport is no longer governed by vacuum tunneling. Instead, the conductance is determined primarily by the geometry of the metallic nanoconstriction. When the transverse dimensions of the contact become comparable to or smaller than the electronic mean free path, transport approaches the ballistic Sharvin regime.\cite{Agrait2003,Scheer1998} The conductance is then proportional to the number of transverse electronic modes crossing the constriction,

\begin{equation}
G \simeq G_0 \frac{k_F^2 A}{4\pi},
\label{eq:Sharvin}
\end{equation}
where \(G_0=2e^2/h\) is the conductance quantum, \(k_F\) is the Fermi wavevector, and \(A\) is the minimum cross-sectional area of the metallic neck.
Assuming an approximately axisymmetric constriction of radius \(r\),
$A=\pi r^2$
so that Eq.~(\ref{eq:Sharvin}) becomes

\begin{equation}
G = G_0\frac{k_F^2 r^2}{4}.
\label{eq:SharvinRadius}
\end{equation}
Introducing the dimensionless conductance
$g={G}/{G_0}$,
one obtains
\begin{equation}
r=\frac{2}{k_F}\sqrt{g}.
\label{eq:rFromG}
\end{equation}
Thus, the experimentally measured quantity $\sqrt{G}$ is directly proportional to the effective neck radius. The conductance traces reveal an approximately linear dependence of $\sqrt{g}$ on time,
\begin{equation}
\sqrt{g(t)}=\sqrt{g_0}+bt,
\label{eq:sqrtglinear}
\end{equation}
which implies
$r(t)=r_0+v_r t$,
with
$v_r={2b}/{k_F}$.
The quantity \(v_r\), therefore, represents the experimentally determined radial growth velocity of the metallic bridge. The approximately constant value of \(v_r\) suggests that once a metallic bridge forms, neck growth is governed by a drift process rather than by purely diffusive transport. A natural mechanism is electromigration-assisted atomic motion driven by the large current densities concentrated within the nanoconstriction.\cite{Sorbello1997,Ho1989,Trouwborst2008}
\begin{figure}
    \centering
    \includegraphics[width=1.1\linewidth]{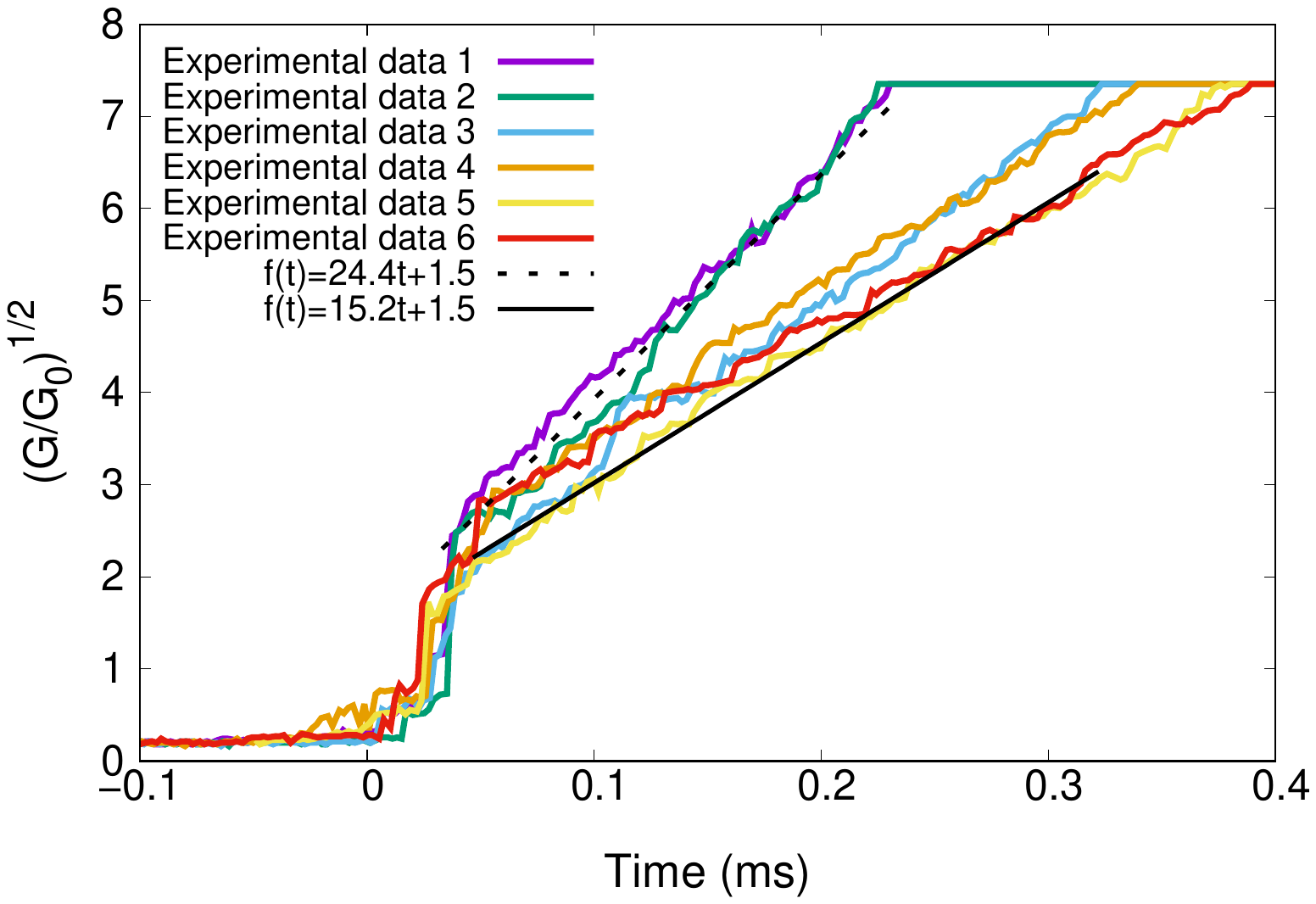}
    \caption{Graph of the conductance dynamics of a metallic nanobridge under green illumination. The approximately linear increase of $\sqrt{G/G_0}$
 indicates drift-like neck growth. The fitted slopes are 24.4 ms$^{-1}$ and 15.2 ms$^{-1}$, corresponding to radial growth velocities $v_r\sim$ 4.1 $\mu$ m/s {and} 2.5 $\mu$m/s.}
    \label{fig:opening-blunting}
\end{figure}

The drift velocity of Au atoms may be written as
$v_{\mathrm{atom}}
=
\mu_a F_{\mathrm{em}}$,
where \(\mu_a\) is the atomic mobility and \(F_{\mathrm{em}}\) is the effective electromigration force. The atomic mobility is thermally activated,
with an activation energy $E_a$ for atomic migration. The experimentally observed radial growth velocity is related to the nanoscopic atomic drift velocity through mass conservation. If atoms are supplied to the neck from a feeding region of effective length \(\ell\), one obtains

\begin{equation}
v_r
=
\frac{\Omega n_s P_{\mathrm{feed}}}
     {2\pi r\,\ell}
\,v_{\mathrm{atom}},
\label{eq:vrvatom}
\end{equation}
where \(\Omega\) is the atomic volume, \(n_s\) is the surface atomic density, and \(P_{\mathrm{feed}}\) is the effective perimeter feeding the neck. For an approximately cylindrical geometry, \(P_{\mathrm{feed}}\simeq 2\pi r\), yielding
\begin{equation}
v_r
=
\frac{\Omega n_s}{\ell}
\,v_{\mathrm{atom}}.
\label{eq:vrvatom2}
\end{equation}
Since \(\Omega n_s\) is of the order of an atomic length, Eq.~(\ref{eq:vrvatom2}) provides the connection between the microscopic atomic drift and the experimentally measured growth velocity.

Figure~\ref{fig:opening-blunting} shows the plot of $\sqrt{g}$ as a function of time under green illumination after bridge formation, which satisfies Eq.~\ref{eq:sqrtglinear} beyond the tunneling regime.  Repeated switching cycles reveal a systematic reduction of the slope \(b\) in Eq.~(\ref{eq:sqrtglinear}), followed by an apparent saturation. We interpret this behavior as a consequence of progressive blunting of the Au asperities that initially define the nanocontact geometry. As the contact becomes smoother, both the local current density and any residual field enhancement decrease, reducing the effective electromigration force \(F_{\mathrm{em}}\) and, therefore, the growth velocity \(v_r\).

From the figure we can compute ${v}_r=2/k_F (d\sqrt{g}/dt)$, using $k_F=1.2\times 10^{10}{\rm m}^{-1}$, then $v_r^{\rm fast}=4.1\mu {\rm m}/{\rm s}$, while $v_r^{\rm slow}=2.5\mu {\rm m}/{\rm s}$. Using our relation to atomic velocities, assuming a feeding region with $\ell\sim 1~{\rm nm}$, we obtain $10-16~\mu {\rm m}/s$. This is much larger than the $1 ~{\rm nm/s}$ reported in ref.\cite{Singh}, but our current densities at the contact can be estimated to be $j\sim 10^{14}$ A/$m^2$ due to crowding in a very small neck, as in the geometry of a breakjunction. We also have an atom-surface flow rather than a bulk flow, as atoms move from the collecting region to the neck. Reference \cite{Trouwborst2008} does not report specific numbers for a similar geometry.
\begin{figure}
    \centering
    \includegraphics[width=1.1\linewidth]{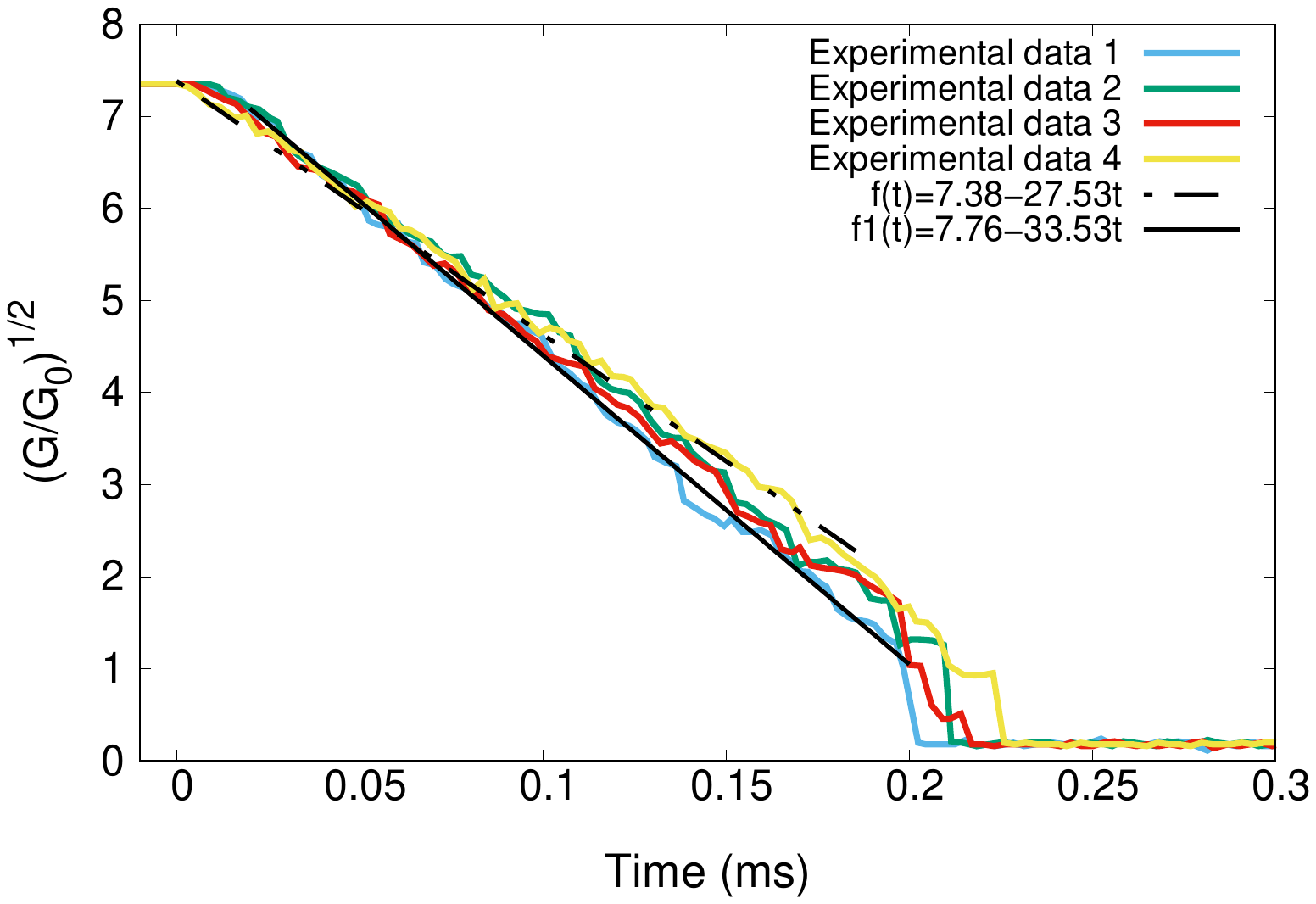}
    \caption{Graph of the conductance dynamics of a metallic nanobridge under blue illumination. The approximately linear decrease of $\sqrt{G/G_0}$ indicates drift-like neck thinning. The fitted slopes are $-27.53~{\rm ms}^{-1}$ and $-33.53~{\rm ms}^{-1}$, corresponding to radial thinning velocities $|v_r|\simeq 4.6~\mu{\rm m/s}$ and $5.6~\mu{\rm m/s}$, respectively.}
    \label{fig:openingContactSharvin}
\end{figure}
Using Eq.~(\ref{eq:rFromG}), conductances in the range $G\sim 50\text{--}100\,G_0$ correspond to effective neck radii of approximately
$r\sim 1\text{--}2~\mathrm{nm}$ consistent with metallic Au nanocontacts reported in the break-junction literature.

We can use the same theory discussed before to analyze the results for rupturing contact depicted in Fig.~\ref{fig:openingContactSharvin}. Using the same parameters, we get radial velocities of $v_r=4.6-5.6\mu {\rm m}/{\rm s}$, and $v_{\rm atom}=19-23\mu {\rm m}/{\rm s}$ for $\ell\sim 1$nm, which is slightly faster than the green-light neck-growth speed but constitutes the same process of electromigration. A minor change of slope as a function of cycling may be observed, but it is not completely systematic.

\begin{figure}
    \centering
    \includegraphics[width=1.1\linewidth]{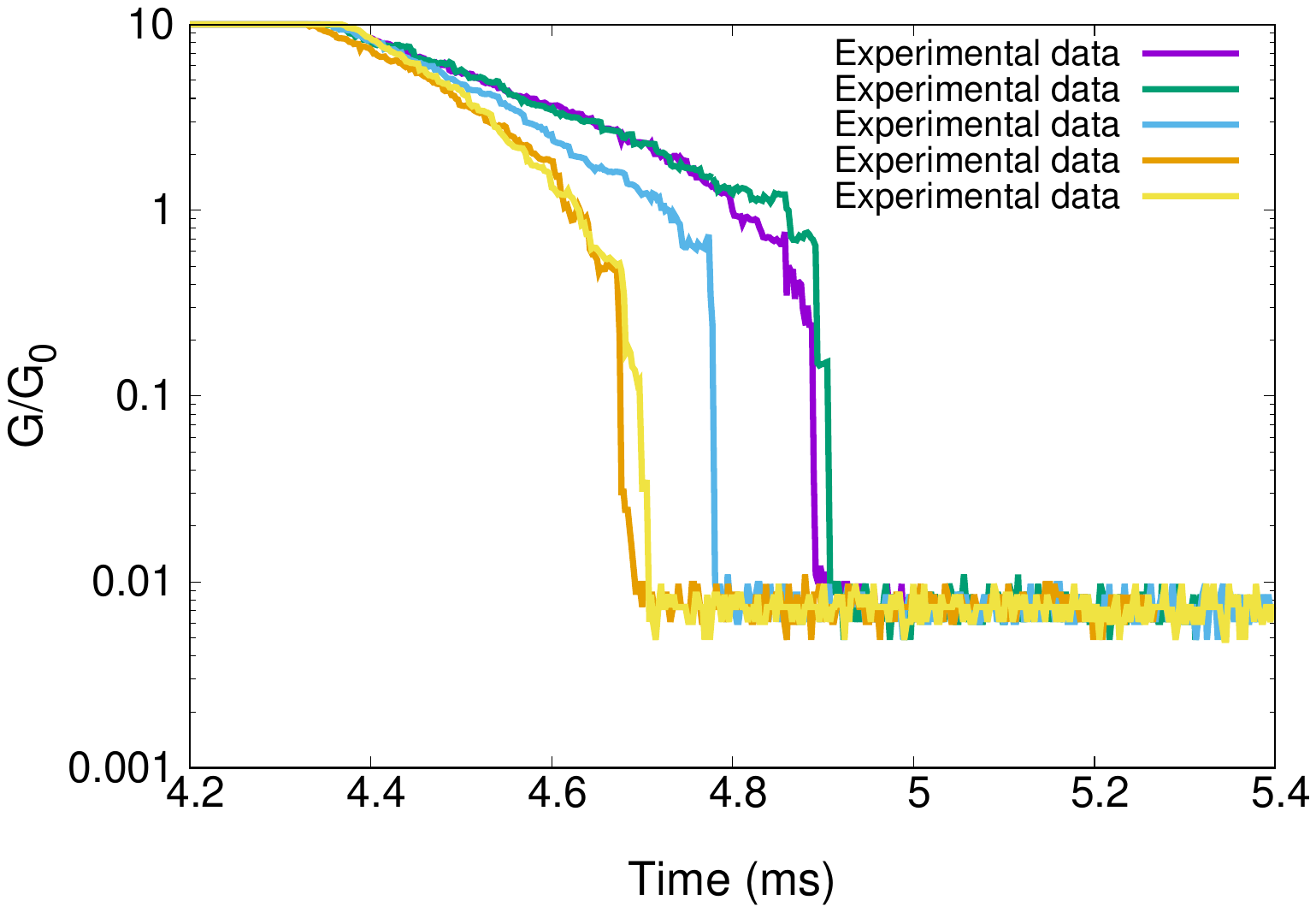}
    \caption{Dark-bridge rupture after green illumination is removed. The rupture trajectory is less systematic than in the illuminated case.}
    \label{fig:Construction-log-oscuro}
\end{figure}

\begin{figure}
    \centering
    \includegraphics[width=1.1\linewidth]{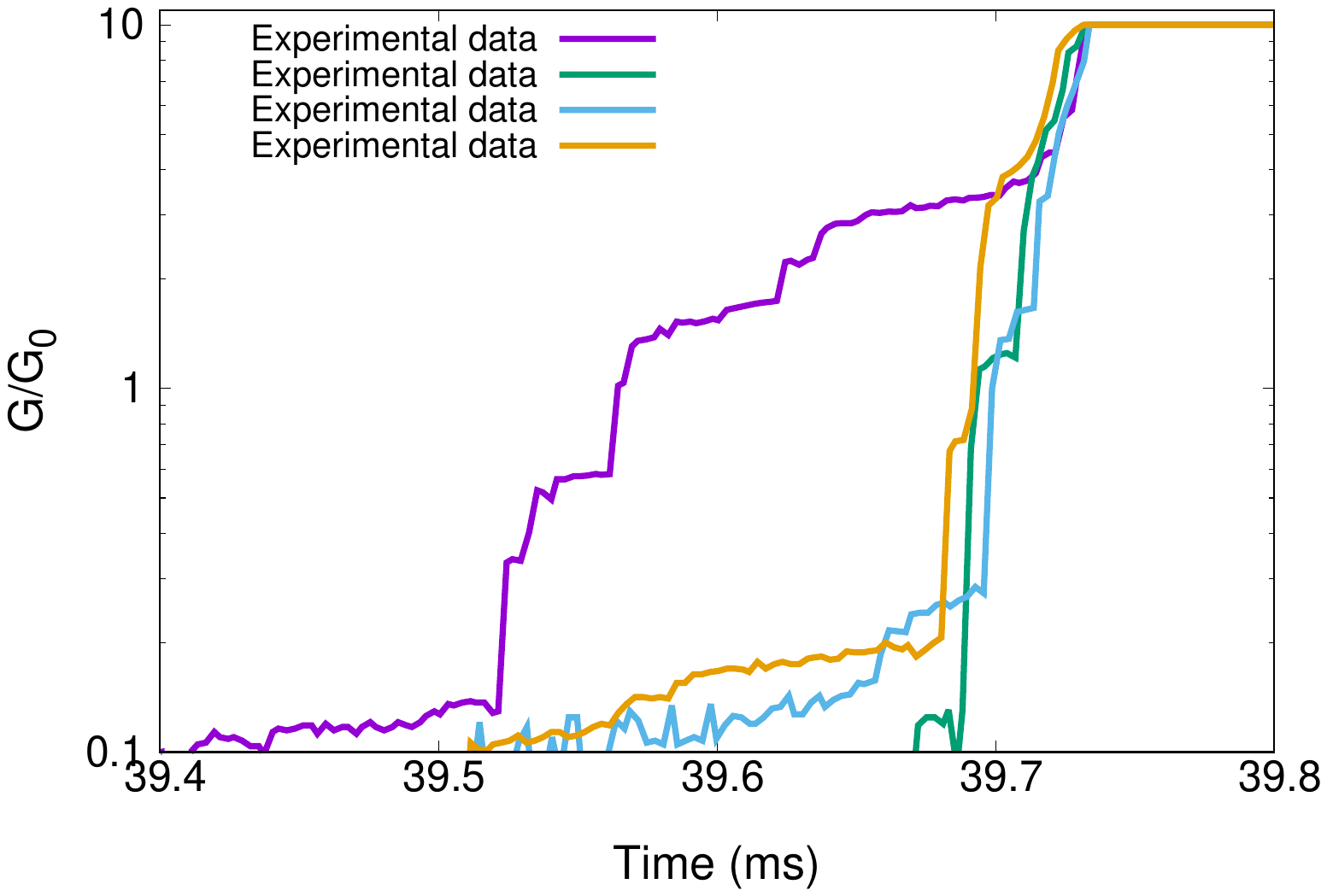}
    \caption{Dark bridge formation after blue illumination is removed. The contact follows different trajectories toward bridge formation.}
    \label{fig:Destrucion-log-oscuro}
\end{figure}


\subsection{Dark bridge rupture and dark bridge formation after illumination}

Both light-prepared states remain stable while the corresponding laser is on. When the light is removed, the junction relaxes back toward the mechanically preferred configuration. After green-induced bridge formation, the bridge ruptures in the dark, as shown in Fig.~\ref{fig:Construction-log-oscuro}. Conversely, after blue-induced bridge rupture, the bridge reforms in the dark, as shown in Fig.~\ref{fig:Destrucion-log-oscuro}. These dark trajectories are less systematic than the light-driven ones, indicating that illumination supplies a wavelength-dependent driving term rather than simply setting a permanent junction geometry. 

\section{Wavelength-Dependent Effective Energy Landscape}


The opposite signs of the optically induced atomic motion suggest that illumination does not act merely as a scalar heat source. Instead, it may modify the effective stability of competing atomic configurations of the junction. To describe this phenomenologically, we introduce a collective coordinate $q$ representing the junction geometry, such as the effective electrode separation, the neck radius, or the number of atoms forming the metallic bridge.

In the absence of illumination, the contact is characterized by a free-energy landscape $F_0(q)$ containing metastable minima associated with a bridged metallic configuration, denoted by $q_{\rm B}$, and a ruptured or tunneling configuration, denoted by $q_{\rm R}$. Under optical excitation, we write the effective landscape as
\begin{equation}
F_{\rm eff}(q;\lambda,I)
=
F_0(q)
+
\Delta F_{\rm opt}(q;\lambda,I),
\label{eq:effective_free_energy}
\end{equation}
where $\lambda$ and $I$ are the illumination wavelength and intensity, respectively. 
The term $\Delta F_{\rm opt}$ represents, at a phenomenological level, geometry-dependent modifications of the activation landscape
arising from plasmonic near fields, hot-carrier populations, optical forces, interface dipoles, and possible adsorbate-mediated effects. Photothermal enhancement of the transition rates is incorporated
primarily through the local effective temperature $T_{\rm eff}$.


Because both the optical near field and the absorbed power depend sensitively on the atomic geometry, $\Delta F_{\rm opt}$ need not be the same for the bridged and ruptured configurations. Illumination can therefore modify both the relative stability of the two states and the activation barriers separating them. The two relevant activated processes are bridge formation, $q_{\rm R}
\longrightarrow
q_{\rm B}$, and bridge rupture, $q_{\rm B}
\longrightarrow
q_{\rm R}$.
Let $q_{\rm R\rightarrow B}^{\ddagger}$ and
$q_{\rm B\rightarrow R}^{\ddagger}$ denote the corresponding transition-state configurations. 
This construction is conceptually related to nonequilibrium descriptions of current-induced barrier modification in atomic
contacts~\cite{ZhangRungger2011}. In the present case, the corresponding nonequilibrium correction is introduced phenomenologically as an optically induced modification of the activation landscape.


In terms of effective activation barriers,
\begin{equation}
\Delta F_{\rm R\rightarrow B}^{\ddagger}(\lambda,I)
=
F_{\rm eff}
\left(
q_{\rm R\rightarrow B}^{\ddagger};
\lambda,I
\right)
-
F_{\rm eff}
\left(
q_{\rm R};
\lambda,I
\right),
\label{eq:formation_barrier}
\end{equation}
and
\begin{equation}
\Delta F_{\rm B\rightarrow R}^{\ddagger}(\lambda,I)
=
F_{\rm eff}
\left(
q_{\rm B\rightarrow R}^{\ddagger};
\lambda,I
\right)
-
F_{\rm eff}
\left(
q_{\rm B};
\lambda,I
\right).
\label{eq:rupture_barrier}
\end{equation}
so the corresponding transition rates may be written as
\begin{equation}
k_{\rm R\rightarrow B}(\lambda,I)
=
\nu_{\rm R\rightarrow B}
\exp
\left[
-
\frac{
\Delta F_{\rm R\rightarrow B}^{\ddagger}(\lambda,I)
}{
k_{\rm B}T_{\rm eff}(\lambda,I)
}
\right],
\label{eq:formation_rate}
\end{equation}
and
\begin{equation}
k_{\rm B\rightarrow R}(\lambda,I)
=
\nu_{\rm B\rightarrow R}
\exp
\left[
-
\frac{
\Delta F_{\rm B\rightarrow R}^{\ddagger}(\lambda,I)
}{
k_{\rm B}T_{\rm eff}(\lambda,I)
}
\right],
\label{eq:rupture_rate}
\end{equation}
where $\nu_{\rm R\rightarrow B}$ and
$\nu_{\rm B\rightarrow R}$ are effective attempt frequencies, and
$T_{\rm eff}$ is the local effective temperature of the active junction region.
The direction of the observed atomic evolution is determined by the imbalance between these two rates. Under the formation wavelength,
$\lambda_{\rm form}\simeq 530\,{\rm nm}$, the optical excitation may lower the barrier for bridge formation, stabilize the bridged configuration relative to the ruptured configuration, or produce both effects. In this case,
\begin{equation}
k_{\rm R\rightarrow B}
\left(
\lambda_{\rm form},I
\right)
>
k_{\rm B\rightarrow R}
\left(
\lambda_{\rm form},I
\right),
\label{eq:formation_bias}
\end{equation}
and the junction evolves toward gap closure and the formation of a metallic bridge.

Conversely, under the rupture wavelength,
$\lambda_{\rm rup}\simeq 407\,{\rm nm}$, the optical excitation may lower the rupture barrier or stabilize the open configuration, leading to
\begin{equation}
k_{\rm B\rightarrow R}
\left(
\lambda_{\rm rup},I
\right)
>
k_{\rm R\rightarrow B}
\left(
\lambda_{\rm rup},I
\right).
\label{eq:rupture_bias}
\end{equation}
The junction then evolves toward neck thinning, bridge rupture, and subsequent gap opening.

This picture naturally accounts for the observation that different wavelengths bias the atomic dynamics in opposite directions. A purely thermal mechanism would primarily increase atomic mobility and accelerate transitions between available configurations. By contrast, the observed reversal in the direction of atomic motion suggests that the optical excitation also modifies the relative activation barriers or effective stability of the competing junction geometries. Illumination therefore acts not only by increasing the local temperature but also by reshaping the effective landscape that governs atomic contact.

The effective-free-energy description should be understood as a phenomenological extension of the plasmonic-heating mechanism proposed by Zhang \emph{et al.}~\cite{Zhang2019}. In their interpretation, plasmon-induced heating induces thermal expansion of the electrodes, thereby controlling the junction conductance. In the present picture, the optical excitation may additionally modify the relative stability and transition barriers of the bridged and ruptured atomic configurations. Since the illuminated junction is a driven nonequilibrium system, $F_{\rm eff}$ should not necessarily be interpreted as a strict equilibrium thermodynamic free energy, but rather as an effective potential governing the activated atomic dynamics.

\section{Discussion and Conclusions}

We have demonstrated wavelength-selective atomic motion in a
mechanically controllable Au break junction. At mechanically fixed
operating points, illumination at
$\lambda_{\rm form}\simeq 530\,{\rm nm}$ drives gap closure and
metallic bridge formation, whereas illumination at
$\lambda_{\rm rup}\simeq 407\,{\rm nm}$ drives neck thinning,
bridge rupture, and subsequent gap opening. The reproducibility of
these processes over repeated cycles shows that illumination supplies
a wavelength-dependent driving term for the atomic dynamics.

The conductance transients reveal two complementary dynamical regimes.
In the tunneling regime, $G<G_0$, the conductance follows
\begin{equation}
G(d)=G_{\rm pref}\exp(-\beta d),
\end{equation}
so an exponential dependence of $G$ on time corresponds to an
approximately constant velocity of the gap coordinate. For effective
barrier heights in the range $\phi\simeq 3$--$5\,{\rm eV}$, the
formation trace gives a gap-closing velocity
$v_f\simeq 2.4$--$3.1\,\mu{\rm m\,s^{-1}}$, whereas the rupture trace
gives a gap-opening velocity
$v_b\simeq 5.2$--$6.8\,\mu{\rm m\,s^{-1}}$. Since
$v\propto\phi^{-1/2}$, varying the effective barrier across this full
range changes the inferred velocities by only about $30\%$ and does
not affect the qualitative conclusions.

In the metallic regime, $G>G_0$, the Sharvin relation gives
\begin{equation}
r=\frac{2}{k_F}\sqrt{\frac{G}{G_0}},
\end{equation}
so the observed linear dependence of $\sqrt{G/G_0}$ on time implies
an approximately constant radial velocity of the metallic neck.
Under green illumination, the bridge grows with
$v_r\simeq 2.5$--$4.1\,\mu{\rm m\,s^{-1}}$, whereas under blue
illumination it thins with
$|v_r|\simeq 4.6$--$5.6\,\mu{\rm m\,s^{-1}}$. The close agreement
between the blue-light thinning velocity in the metallic regime and
the gap-opening velocity in the tunneling regime supports a continuous
rupture process: the neck first thins and, after contact is lost, the
The newly formed gap continues to widen at a comparable rate.

The reversal of the atomic motion with wavelength cannot be described
as a simple consequence of uniform heating. We interpret it
phenomenologically as a wavelength-dependent modification of the
effective activation landscape of the contact. In the nanogap regime,
different optical excitations may favor attractive or destabilizing
gap configurations. Once a metallic bridge is formed, the capacitive
gap mode is quenched, and the subsequent dynamics are more naturally
associated with current crowding, enhanced surface mobility, local
heating, and electromigration-assisted atomic motion. The dark
relaxation observed after the illumination is removed further shows
that the optically driven state competes with the mechanical bias of
the junction rather than representing a permanent structural change.

These results provide a quantitative description of optically selected
atomic drift across the tunneling and metallic-contact regimes. They
show that the direction of atomic rearrangement in Au nanocontacts can
be selected by the illumination wavelength and followed directly
through time-resolved conductance measurements.

\acknowledgments
 DL and EM acknowledge financial support from the Universidad San Francisco de Quito through Poligrant POLI~41981. 
 CS gratefully acknowledges financial support from the Generalitat Valenciana (CIDEXG/2022/45) and the Spanish Government through  PID2023-146660OB-I00, received funding from MICIU/AEI/10.13039/501100011033, and the European Regional Development Fund (ERDF/EU). We are grateful to Oren Tal's group for guiding us in setting up our break junction device, and for many fruitful ideas in making it a precision-level instrument.

\section*{Declarations}

\subsection*{Conflict of interest}

The authors declare that they have no known competing financial interests or personal relationships that could have appeared to influence the work reported in this article.

\subsection*{Data availability statement}

The data that support the findings of this study are available from the corresponding author upon reasonable request.

\subsection*{Ethics statement}

This study did not involve human participants, human data, or animals. Therefore, ethical approval was not required.

\newpage

\bibliography{bibliography}

@article{HoffmannVogel2017,
  author  = {Hoffmann-Vogel, R.},
  title   = {Electromigration and the structure of metallic nanocontacts},
  journal = {Appl. Phys. Rev.},
  volume  = {4},
  pages   = {031302},
  year    = {2017}
}

@article{ward2010l,
  title = {Optical rectification and field enhancement in a plasmonic nanogap},
  author = {Ward, Daniel R. and H{\"u}ser, F. and Pauly, F. and Cuevas, Juan Carlos and Natelson, Douglas},
  journal = {Nature Nanotechnology},
  volume = {5},
  number = {10},
  pages = {732--736},
  year = {2010},
  publisher = {Nature Publishing Group}
}

@article{herzog2013dark,
  title = {Dark plasmons in hot spot generation and polarization in interelectrode nanoscale junctions},
  author = {Herzog, Joseph B. and Knight, Mark W. and Li, Yajing and Evans, Kenneth M. and Halas, Naomi J. and Natelson, Douglas},
  journal = {Nano Letters},
  volume = {13},
  number = {3},
  pages = {1359--1364},
  year = {2013},
  publisher = {ACS Publications}
}

@article{benz2016,
  title = {Single-molecule optomechanics in "picocavities"},
  author = {Benz, Felix and Schmidt, Mikolaj K. and Dreismann, Alexander and Chikkaraddy, Rohit and Zhang, Yao and Demetriadou, Angela and Carnegie, Cloudy and Ohadi, Hamid and de Nijs, Bart and Esteban, Rub{\'e}n and Aizpurua, Javier and Baumberg, Jeremy J.},
  journal = {Science},
  volume = {354},
  number = {6313},
  pages = {726--729},
  year = {2016},
  publisher = {American Association for the Advancement of Science}
}

@book{Cuevasbook,
  title = {Molecular Electronics},
  author = {Cuevas, Juan Carlos and Scheer, Elke},
  publisher = {World Scientific},
  year = {2017},
  edition = {2nd},
  doi = {10.1142/10598},
  url = {https://www.worldscientific.com/doi/abs/10.1142/10598},
  eprint = {https://www.worldscientific.com/doi/pdf/10.1142/10598}
}

@article{Pascual1993,
  title = {{Quantum contact in gold nanostructures by scanning tunneling microscopy}},
  author = {Pascual, J. I. and M\'endez, J. and G\'omez-Herrero, J. and Bar\'o, A. M. and Garc\'{\i}a, N. and Binh, Vu Thien},
  journal = {Phys. Rev. Lett.},
  volume = {71},
  issue = {12},
  pages = {1852--1855},
  year = {1993},
  month = {Sep},
  publisher = {American Physical Society},
  doi = {10.1103/PhysRevLett.71.1852},
  url = {https://link.aps.org/doi/10.1103/PhysRevLett.71.1852}
}

@article{Krans93,
  title = {One-atom point contacts},
  author = {Krans, J. M. and Muller, C. J. and Yanson, I. K. and Govaert, Th. C. M. and Hesper, R. and van Ruitenbeek, J. M.},
  journal = {Phys. Rev. B},
  volume = {48},
  issue = {19},
  pages = {14721--14724},
  year = {1993},
  month = {Nov},
  publisher = {American Physical Society},
  doi = {10.1103/PhysRevB.48.14721},
  url = {https://link.aps.org/doi/10.1103/PhysRevB.48.14721}
}

@article{Krans96,
  title = {Atomic structure and quantized conductance in metal point contacts},
  author = {Krans, J. M. and van Ruitenbeek, J. M. and de Jongh, L. J.},
  journal = {Physica B: Condens. Matter},
  volume = {218},
  pages = {228},
  year = {1996}
}

@article{Landauer57,
  title = {Spatial Variation of Currents and Fields Due to Localized Scatterers in Metallic Conduction},
  author = {Landauer, R.},
  journal = {IBM Journal of Research and Development},
  volume = {1},
  number = {3},
  pages = {223--231},
  year = {1957},
  doi = {10.1147/rd.13.0223}
}

@article{sharvin1965,
 title = {A possible method for studying Fermi surfaces}, 
author = {Sharvin, Yu. V.},
 journal = {Sov. Phys. JETP}, 
volume = {21}, 
number = {3}, 
pages = {655--656}, year = {1965} 
}

@article{ZhangRungger2011,
  author  = {Zhang, R. and Rungger, I. and Sanvito, S. and Hou, S.},
  title   = {},
  journal = {Phys. Rev. B},
  volume  = {84},
  pages   = {085445},
  year    = {2011}
}

@article{Agrait2003,
  author  = {Agraït, N. and Yeyati, A. L. and van Ruitenbeek, J. M.},
  journal = {Phys. Rep.},
  volume  = {377},
  pages   = {81--279},
  year    = {2003}
}

@article{Scheer1998,
  author  = {Scheer, E. and Agraït, N. and Cuevas, J. C. and Yeyati, A. Levy and Ludoph, B. and Martín-Rodero, A. and Bollinger, G. R. and van Ruitenbeek, J. M. and Urbina, C.},
  journal = {Nature},
  volume  = {394},
  pages   = {154--157},
  year    = {1998}
}

@article{Zhang2019,
  author  = {Zhang, W. and Liu, H. and Lu, J. and Ni, L. and Liu, H. and Li, Q. and Qiu, M. and Xu, B. and Lee, T. and Zhao, Z. and Wang, X. and Wang, M. and Wang, T. and Offenhäusser, A. and Mayer, D. and Hwang, W.-T. and Xiang, D.},
  journal = {Light: Sci. Appl.},
  volume  = {8},
  pages   = {34},
  year    = {2019}
}

@article{Marchesin2015,
  author  = {Marchesin, F. and Koval, P. and Sánchez-Portal, D. and Rubio, A.},
  journal = {Phys. Rev. B},
  volume  = {91},
  pages   = {115431},
  year    = {2015}
}

@article{Khurgin2024,
  author  = {Khurgin, J. B. and others},
  journal = {eLight},
  volume  = {4},
  pages   = {9},
  year    = {2024}
}

@article{Zhou2024,
  author  = {Zhou, Z. L. and Zhu, X. and others},
  journal = {Chem. Rev.},
  volume  = {124},
  pages   = {7131--7184},
  year    = {2024}
}

@incollection{Sorbello1997,
  author    = {Sorbello, R. S.},
  title     = {Theory of Electromigration},
  booktitle = {Solid State Physics},
  editor    = {Ehrenreich, H. and Spaepen, F.},
  publisher = {Academic Press},
  address   = {San Diego},
  volume    = {51},
  pages     = {159--231},
  year      = {1997}
}

@article{Ho1989,
  author  = {Ho, P. S. and Kwok, T.},
  journal = {Rep. Prog. Phys.},
  volume  = {52},
  pages   = {301--348},
  year    = {1989}
}

@article{Trouwborst2008,
  author  = {Trouwborst, M. L. and van der Molen, S. J. and van Wees, B. J.},
  journal = {Appl. Phys. Lett.},
  volume  = {93},
  pages   = {043118},
  year    = {2008}
}

@article{Nordlander2004,
  author  = {Nordlander, P. and Oubre, C. and Prodan, E. and Li, K. and Stockman, M. I.},
  journal = {Nano Lett.},
  volume  = {4},
  pages   = {899--903},
  year    = {2004}
}

@article{Singh,
  author  = {Singh, A. K. and Kumar, J.},
  journal = {J. Vac. Sci. Technol. B},
  volume  = {38},
  pages   = {062805},
  year    = {2020}
}



\appendix

\end{document}